# Artificial Intelligence and Legal Analysis: Implications for Legal Education and the Profession

Lee F. Peoples[*]

---

(forthcoming 117 Law Library Journal ____ (2025)

---

## Abstract

This article reports the results of a study examining the ability of legal and non-legal Large Language Models (LLMs) to perform legal analysis using the Issue-Rule-Application-Conclusion (IRAC) framework. LLMs were tested on legal reasoning tasks involving rule analysis and analogical reasoning. The results show that LLMs can conduct basic IRAC analysis, but are limited by brief responses lacking detail, an inability to commit to answers, false confidence, and hallucinations. The study compares legal and non-legal LLMs, identifies shortcomings, and explores traits that may hinder their ability to "think like a lawyer." It also discusses the implications for legal education and practice, highlighting the need for critical thinking skills in future lawyers and the potential pitfalls of over-reliance on artificial intelligence (AI) resulting in a loss of logic, reasoning, and critical thinking skills.

## Implications for Practice

1. Based on legal analysis exercises, Lexis+ AI, ChatGPT, Claude, Gemini, and Copilot have different legal research abilities and shortcomings.

2. Large Language Models have a propensity to hallucinate, have false confidence in incorrect answers, and are unable to reason based on principles, policies, or moral thinking.

3. LLMs' limitations regarding legal analysis include lack of stable outputs, non-transparency of training corpus and algorithms, constraints on research creativity, and the inability to conform LLM outputs to applicable ethical rules and norms.

4. AI may create future generations of law students lacking in critical thinking, logic, and reasoning abilities.

---

# Introduction

¶1 Law and society at large are in the midst of a fourth industrial revolution brought about by artificial intelligence and related technologies.[1] These developments could potentially usher in a "golden age for the [legal] profession and society"[2] but also have the potential to create "economic inefficiency, social disfunction, and a declining legal system."[3] Goldman Sachs predicted that AI could be more disruptive in law than other industries and could potentially replace 44 percent of legal jobs.[4]

¶2 Previous studies have concluded that law students using generative AI tools could "substantially improve the efficiency with which they complete a broad array of legal tasks without adversely affecting (or even slightly improving) the quality of that work product."[5] Lawyers will also benefit from AI, and it will be a vital tool for law practice both in the future and in the long term.[6]

¶3 However, the gains to be realized from using AI in legal education and the legal profession are not without potential perils. AI research tools have notoriously hallucinated facts and laws, fabricated case law that does not exist,[7] falsely attributed a Supreme Court dissenting opinion to a justice who actually joined the majority,[8] falsely accused a law professor of sexual assault,[9] and even temporarily going insane.[10] Lawyers have misused AI to create fictional cases cited in court filings[11] and to estimate attorney's fees, leading to criticism, sanctions from judges, and disciplinary actions.[12]

¶4 This study examines the abilities of legal and non-legal LLMs to think like a lawyer by performing legal analysis and reasoning on legal factual scenarios using the Issue-Rule-Application-Conclusion (IRAC) framework. LLMs tested include Lexis+ AI, Claude,

---

ChatGPT 3.5, Copilot 365, and Gemini. The legal analysis and reasoning tasks used in this study required LLMs to analyze legal rules and statutes and to perform analogous reasoning. Several prompting strategies were utilized to iteratively improve LLMs' responses. The study results demonstrate that LLMs are capable of basic legal reasoning by analyzing legal fact patterns using the IRAC format. The LLMs evaluated demonstrated various levels of success at performing IRAC analysis when analyzing legal rules, statutes, and reasoning analogically.

¶5 Comparisons are made between the findings of this study and other previously published studies on the ability of LLMs to perform legal analysis and reasoning tasks. The legal reasoning and analysis abilities of legal LLMs are compared with those of non-legal LLMs. Shortcomings of LLMs are identified, including the tendency of LLMs to over-generalize and refuse to commit to a conclusive answer. Some LLMs demonstrated false confidence in incorrect answers.

¶6 Traits of LLMs that have the potential to hinder their ability to think like a lawyer, thereby limiting their usefulness to law students and lawyers, are also explored. A trait of LLMs that is troublesome for lawyers is the lack of consistency in responses to identical prompts.[13] Replicability and stability are the cornerstones of legal information. It is difficult to think like a lawyer when source materials are constantly changing.

¶7 When using LLMs for legal research, it is important for lawyers to know what is and is not included in the LLMs' datasets and what techniques to use to retrieve the most relevant information. The companies who built the LLMs tested in this study are notoriously secretive about the corpora LLMs were trained on and how the algorithms that run LLMs operate. The lack of information about these important details limits the ability of LLMs to perform some of the higher-level components of thinking like a lawyer.

¶8 Linguists and legal theorists have critiqued LLMs. Some doubt LLMs' abilities to engage in moral thinking or to reason based on policies or principles. This study includes examples of LLMs failing to base their reasoning on policies and principles. The inability of LLMs to make judgments about what the law should be or to make moral judgments will keep LLMs and other AI technologies from replacing human lawyers and judges.

¶9 Lawyers, unlike LLMs, are officers of the court and are required to constrain their speech and actions by the rules of legal ethics and other norms. Currently, LLMs have not demonstrated the ability to think like a lawyer while also conforming their responses to the legal professions' professional responsibility rules.

¶10 Law schools should be prepared for the future impacts of machine learning on law students and the legal profession. Legal educators are currently working to improve the technology abilities of a generation of law students. These millennials and other generations were incorrectly presumed to be highly competent with technology simply because they grew up using it.

---

¶11 Legal education may encounter similar hurdles when students currently in primary, secondary, and post-secondary schools enter law school. Education thought leaders warn of the unintended consequences of revamping school curriculum with a heavy focus on AI. Students who are overly reliant on AI may never develop logic, reasoning, and critical thinking skills. These skills are the basic building blocks of thinking like a lawyer.

## How LLMs Work[14]

¶12 Large language models are neural networks with billions of parameters trained on vast datasets of text. At their core, they represent words as high-dimensional numerical vectors. These word vectors are fed into layers of artificial neurons that perform mathematical operations to capture linguistic patterns and contexts.[15] This process is an "algorithmic approach to machine learning known as a 'neural network'"[16] The neural network is an "extremely flexible pattern detector."[17] LLMs using neural networks have been described as "advanced AI word-prediction systems."[18]

¶13 When a user enters a prompt into an LLM, it is processed within the LLM's "context window," which functions as its "short-term memory," allowing it to make an "educated guess" about what word should come next.[19] The context window expands as the LLM guesses and generates more words. The words come from the corpus of data that the LLM has been "pretrained"[20] on. Pretraining involves "teaching an AI model how to understand and generate human-like text by exposing the model to … billions of webpages, books, contracts, legal opinions, and other text documents."[21] Pretraining gives an AI system "billions of parameters" … "appropriately adjusted to reliably predict the next words, given nearly any selection of prompting words."[22]

¶14 LLMs use the additional step of "deep learning" to scale a neural network to include "billions of neurons or parameters with many deep layers."[23] Innovations in transformer architecture have improved the accuracy of LLMs by allowing them "to look at the entire context of the user input, even words that were far away, and determine which contextual words were most helpful in figuring out the more accurate next word."[24]

---

14 This section was adapted from the excellent and detailed explanation of machine learning, AI, and LLMs found in Harry Surden, *ChatGPT, AI Large Language Models, and Law*, 92 FORDHAM L. REV. 1939-1970 (2024).

15 Claude's response to this prompt: "explain how large language models work in one paragraph."

16 Surden, *supra* note 14, 1951.

17 *Id.* at 1961.

18 *Id.*

19 *Id.* at 1952.

20 *Id.* at 1959.

21 *Id.*

22 *Id.* at 1961.

23 *Id.* at 1961

24 *Id.* at 1962, *citing* SINAN OZDEMIR, QUICK START GUIDE TO LARGE LANGUAGE MODELS: STRATEGIES AND BEST PRACTICES FOR USING CHATGPT AND OTHER LLMs 70 (2023).

# Using IRAC to Test Legal Analysis Abilities of LLMs

¶15 This study examines the abilities of legal and non-legal LLMs to think like a lawyer by performing legal analysis and reasoning. A common framework taught in American law schools is IRAC. The IRAC framework breaks legal analysis down into four discrete steps. The IRAC framework can be used to analyze legal rules, statutes, and to reason analogically using caselaw. A helpful explanation is found in the Legal Bench study.

> First, lawyers identify the legal issue in a given set of facts (issue-spotting). An issue is often either (1) a specific unanswered legal question posed by the facts, or (2) an area of law implicated in the facts. Depending on the setting, a lawyer may be told the issue or be required to infer a possible issue.

> Second, lawyers identify the relevant legal rules for this issue (rule-recall). A rule is a statement of law that dictates the conditions that are necessary (or sufficient) for some legal outcome to be achieved. In the United States, rules can come from a variety of sources: the Constitution, federal and state statutes, regulations, and court opinions (case law). Importantly, rules often differ between jurisdictions. Hence, the relevant rule in California might be different than the relevant rule in New York.

> Third, lawyers apply these rules to the facts at hand (rule-application). Application, or the analysis of rule applicability, consists of identifying those facts which are most relevant to the rule, and determining how those facts influence the outcome under the rule. Application can also involve referencing prior cases involving similar rules (i.e., precedent) and using the similarities or differences to those cases to determine the outcome of the current dispute, once pretraining is complete. Finally, lawyers reach a conclusion with regards to their application of law to determine what the legal outcome of those facts are (rule-conclusion).[25]

## *Testing Methodology*

¶16 Factual scenarios were adapted from a text utilized in introductory law school courses on legal research, writing, and legal analysis.[26] Factual scenarios were anonymized by removing specific names to keep from exposing content-specific identifiers to LLMs' corpora. Additionally, LLMs were instructed to not train on the prompts.

¶17 LLMs were tested on a total of seven scenarios. Scenarios requiring beginning and skilled IRAC abilities were selected for testing LLMs abilities to analyze legal rules and perform analogical reasoning.[27] Scenarios requiring beginning, intermediate, and skilled IRAC abilities were selected for testing LLMs' abilities to analyze statutes.[28] LLMs' statutory reasoning abilities were explored in more detail because previous studies have demonstrated

---

LLMs' tendency to hallucinate when analyzing statutes.[29] A separate study showed that GPT-3 has "imperfect prior knowledge"[30] of statutes in its training data, "performs poorly at answering straightforward questions about synthetic statutes"[31] created for testing purposes, and has a "tendency to mistakenly refer to the wrong part of a statute … even when we include the text of the statute in the prompt itself."[32]

¶18 The LLM products evaluated include Lexis+ AI, Anthropic's Claude 3 Sonnet, Open AI's GPT 3.5, Microsoft's Copilot 365, and Google's Gemini lightweight LaMDA.[33] Each model has its own strengths and limitations. At the time of evaluations, Lexis+ AI's content included "Cases, Statutory Codes, Practical Guidance, and our proprietary Mathew Bender collection."[34] Lexis+ AI was the only LLM tested that was specifically designed for legal research.[35] At the time of evaluations, none of the LLMs were directly connected to the internet according to their FAQ documents. The non-legal LLMs were trained on a much larger and more diverse dataset than Lexis+ AI.

¶19 Lexis+ AI differs from the non-legal AIs tested in another important way. Lexis+ AI uses "an extensively enhanced LexisNexis proprietary Retrieval Augmented Generation 2.0 (RAG 2.0) platform.[36] RAG works by "examining a user query"[37] and "reach[ing] out to reliable databases"[38] to "gather documents likely to be relevant to the prompt"[39] and "augment[s] the prompt on behalf of the user.[40] RAG has the potential to produce more reliable results when compared with LLMs not using RAG.[41]

¶20 Unlike other LLMs tested, Lexis+ AI offers the user a choice of specifically prompting the system to "ask a legal question," "generate a draft," or "summarize a case." All queries used in this study were entered into Lexis+ AI as "ask a legal question" except for the beginning statutory exercise asking students to "draft the argument portion of your response to the judge."[42] The option to generate a draft was selected for this exercise.

---

¶21 LLMs were instructed to respond with a temperature of 0.0. In the context of LLMs, setting the requested temperature of a response "controls the randomness of the model's output."[43] Selecting a lower temperature like 0.0 "makes the output of the LLM more deterministic, thus favoring the most likely predictions."[44] Selecting a higher temperature "makes the output more random … creative … but also introduce[s] hallucination."[45]

¶22 Nucleus sampling, sometimes referred to as "top P," was set at 1.0. Nucleus sampling "sets the threshold probability and selects the top tokens whose cumulative probability exceeds the threshold."[46] The temperature setting of 0.0 and top P setting of 1.0 were utilized in a previous study testing GPT-3's ability to perform statutory reasoning.[47] These settings are expected to "maximize reproducibility and minimize hallucination and wandering off topic."[48]

¶23 All prompting was input as a "zero-shot" and no additional examples or demonstrations were provided. In the parlance of LLMs, zero-shot refers to providing no additional examples or demonstrations. Few-shot prompting involves providing an LLM with a few examples of the response expected.[49] Previous studies have demonstrated that LLMs perform better when few-shot prompting (as opposed to zero-shot prompting) is used.[50] However, zero-shot prompting is more realistic and appropriate for the purposes of this study, which models the behavior and abilities of a first-year law student or a legal analysis novice. First-year students could not realistically be expected to provide an LLM with correct and relevant examples of a legal analysis answer while they are in the process of developing legal analysis skills.

¶24 Several iterative prompting strategies were employed to improve LLMs' responses. Iterative prompts were used to get an LLM to consider cases or other items provided in the initial prompt but not mentioned in an LLM's initial response. The iterative prompt of instructing the LLM to think step by step when responding to a query was used. This prompting strategy is sometimes referred to as chain-of-thought reasoning. Previous studies have demonstrated the chain-of-thought prompt to be effective when performing legal analysis and other tasks using LLMs.[51]

---

¶25 Each LLM was provided an iterative prompt to respond to the query with an answer in IRAC format. Lexis+ AI ignored this prompt and provided answers in a narrative format lacking the IRAC structure. All other LLMs (Claude, GPT, Copilot, and Gemini) complied with the iterative prompt to respond to the queries with an answer in IRAC format.

## *Scoring*

¶26 A scoring guide was developed to evaluate LLMs answers to the seven legal analysis and reasoning exercises. LLMs were evaluated on how well they completed six discrete legal analysis tasks. Two additional metrics were used to increase or decrease an LLM's score based on how it responded to an iterative prompt and to account for any hallucinations in responses. An LLM that performed perfectly on all seven problems scored a total of 100 points. Appendix 1 reports individual performance data on each LLM for the seven exercises. An LLM whose answer improved after being prompted to use chain of thought reasoning on all seven scenarios received a perfect chain of thought reasoning score of 8.

1. **Relied on Sources as Instructed**. Zero points were awarded to LLMs that ignored instructions. Some points (1.3) were awarded to LLMs that partially followed instructions. Full points (2) were awarded to LLMs that completely followed instructions.

2. **Issue Identification**. Zero points were awarded to LLMs providing an incorrect issue statement. Some points (1.3) were awarded to LLMs that partially identified the issue. Full points (2) were awarded to LLMs correctly identifying the issue.

3. **Stating the Rule**. Zero points were awarded to LLMs misstating the rule(s). Some points (1.3) were awarded to LLMs partially stating the rule. Full points (2) were awarded to LLMs correctly stating the rule.

4. **Applying the Rule**. Zero points were awarded to LLMs that merely re-stated the rule or incorrectly applied the rule. Some points (1.3) were awarded to LLMs making some correct application of the rule. Full points (2) were awarded to LLMs that completely and correctly applied the rule and included all necessary inferences.

5. **Reaching the Correct Conclusion**. Zero points were awarded to LLMs that re-stated the rule or incorrectly applied a rule. Some points (1.3) were awarded to LLMs making some correct application of a rule. Full points (2) were awarded to LLMs that completely and correctly applied the rule and included all necessary inferences.

6. **Conclusion Stated with Certainty**. Zero points were awarded to LLMs that stated the conclusion without certainty. Some points (1.3) were awarded to LLMs that waivered, hedged, qualified, or equivocated when stating the conclusion. Full points (2) were awarded to LLMs that stated the conclusion with certainty, including any legitimate qualifications.

7. **Correctly Responded to the Prompt to Use Chain of Thought Reasoning**. Zero points were awarded to an LLM that did not change its answer after being prompted to use chain of thought reasoning. 1.143 points were awarded to an LLM whose answer improved after being prompted to use chain of thought reasoning.

8. **Hallucination**. What qualifies as a hallucination was borrowed from the Large Legal Fictions study identifying three distinct types of hallucinations.[52] The first type is a hallucination of the law or facts provided that might "produce a response that is unfaithful to or in conflict with the input prompt."[53] The second type is a response that "contradicts or does not directly derive from its training corpus."[54] The third type is a response that is not consistent "with the facts of the world" including hallucinated statements of law. This study adopts only the first type (hallucination of law or facts provided) and third type (hallucination of the facts of the world or law). The corpora of the non-law LLMs evaluated in this study are vast and only vaguely defined, making it impossible to determine what is and is not included. Zero points were awarded to a hallucinating LLM. A total of 1.143 points were awarded to an LLM that did not hallucinate. An LLM that did not hallucinate on any of the seven scenarios received a perfect hallucination score of 8.

Table 1. LLM Performance
All Exercises Combined

|  | Lexis + AI | Claude | Copilot | GPT 3.5 | Gemini |
|---|---|---|---|---|---|
| **Relied on Sources as Instructed** | 10.500 | 12.600 | 12.000 | 11.900 | 11.200 |
| **Issue Identification** | 11.200 | 13.300 | 12.600 | 11.900 | 11.200 |
| **Stating the Rule** | 11.200 | 12.600 | 12.600 | 12.600 | 10.600 |
| **Applying the Rule** | 7.900 | 12.600 | 9.200 | 8.500 | 8.500 |
| **Reaching Correct Conclusion** | 10.600 | 12.000 | 12.000 | 10.000 | 11.300 |
| **Conclusion Stated with Certainty** | 11.200 | 13.300 | 11.200 | 11.900 | 11.200 |
| **Chain of Thought Prompt** | 3.429 | 6.858 | 6.858 | 4.572 | 5.715 |
| **Hallucination** | 3.429 | 8.001 | 8.001 | 6.858 | 6.858 |
|  |  |  |  |  |  |
| **TOTAL SCORE ___ / 100** | 69.46 | 91.26 | 84.46 | 78.23 | 76.57 |

# Legal Analysis Results
## *Beginning Rule Analysis*

¶27 The beginning rules analysis problem involved the Americans with Disabilities Act[55] and an unhoused person seeking to keep his animal with him in a homeless shelter.[56] The facts include a description of the unhoused person's afflictions of anxiety and depression and an explanation of how the animal provides comfort to the unhoused person, including potentially preventing him from overdosing by removing pills from his hand.

¶28 The problem set out the following rules to be applied to the factual situation:

---

52 Matthew Dahl, et al., *Large Legal Fictions: Profiling Legal Hallucinations in Large Language Models*, arXiv, April 25, 2024, at 9, https://arxiv.org/pdf/2401.01301 [https://perma.cc/BVJ9-T2BC].

53 *Id.*

54 *Id.* at 10.

55 Americans With Disabilities Act of 1990, 42 U.S.C. § 12101 et seq.

56 The beginning rule analysis problem was adapted from exercise 11 in HILL & VUKADIN, *supra* note 26, at 22.

The ADA requires public entities to make reasonable modifications to operations or policies to avoid discrimination based on disabilities. Upon a showing that a modification would "fundamentally alter the nature of the services, program, or activity," a public entity may not be required to make the requested modification.[57] Animals providing comfort to a disabled person are not automatically classified as service animals. The ADA defines service animals as an animal "individually trained to do work or perform similar tasks for the benefit of an individual with a disability."[58] The problem does not state that the animal in question has had any training or special abilities as required under the ADA. Service animals are required to possess abilities beyond those of dogs in general or typical to their breed.

¶29 The correct answer to this problem should conclude that the ADA would not recognize the animal as a "service animal" because it lacks the specific training required for classification as a service animal. Ultimately, the unhoused individual will not be able to require the shelter to accommodate his animal according to the ADA.

### LLMs' Responses – Beginning Rule Analysis

¶30 Copilot outperformed other LLMs at analyzing the beginning rule analysis question. Copilot's answer contained significantly more relevant details than some of the other models. The answer correctly identified the key issue of the animal's lack of training as required by the ADA. Claude performed well but its responses were shorter and lacked the detail provided by Copilot's response.

¶31 GPT and Gemini both hallucinated the facts provided and made the incorrect assumption that the animal's act of potentially preventing its owner from overdosing by removing pills from his hand could meet the ADA requirement of having special training or abilities as a service animal. Both hedged their bets on the final conclusion and qualified their answers that more investigation is needed to determine if the animal qualifies as a service animal under the ADA.

¶32 Lexis+ AI did not respond in a meaningful way to the iterative prompts to answer thinking step by step or to respond in IRAC format. Lexis+ AI ignored or was not able to process the instructions for any of the six discrete legal analysis tasks. Lexis+ AI included citations to relevant ADA regulations found in the Code of Federal Regulations (CFR). Lexis+ AI failed to identify the issue that the animal does not qualify as a service animal because it has not received the training required under the ADA. Ultimately, Lexis+ AI's conclusion was to name the animal in question and then re-state the ADA language defining a service animal as "any guide dog, signal dog, or other animal individually trained to provide assistance to an individual with a disability." Lexis+ AI ignored the premise in the facts that the dog had not received any special training. Lexis+ AI's reply included a hallucination of the facts provided by responding with information about miniature horses and non-human primates as service animals under the ADA.

---

57  28 C.F.R. § 35.130(b)(7) (2024).

58  28 C.F.R. § 36.104 (2024).

## *Skilled Rule Analysis*

¶33 The skilled rules analysis problem involves questions of criminal law. Law enforcement is seeking a 30-day search warrant for continuous and detailed location data on a suspect from the suspect's cellular phone service provider.[59] The search warrant is sought to aid law enforcement in executing an arrest warrant on the suspect. No claim has been made that the location data will provide evidence of a crime. The problem sets out the following rules to be applied to the factual situation:

- Under the Fourth Amendment, subjects of an arrest warrant have reasonable expectations of privacy in their movements and location.

- Search warrants can be justified under the Fourth Amendment by showing the evidence to be obtained has some nexus with criminal activity.

- The correct answer to this problem should evaluate whether an arrest warrant alone provides a legal basis to obtain continuous and detailed location data on a suspect for 30 days.

### LLMs' Responses - Skilled Rule Analysis

¶34 Claude's answer was the most complete and thorough of the group, its conclusion was stated with certainty, and citations to relevant federal criminal law cases were included with brief comments explaining the relevance of the cases to the fact pattern.

¶35 Lexis+ AI provided a solid answer that included links to relevant case law within Lexis. However, its ultimate answer lacked the certainty found in answers returned by other LLMs.

¶36 Copilot's performance was the worst of all LLMs evaluated at skilled rule analysis. Copilot failed to clearly and distinctly identify the two rules stated in the problem. Copilot's initial answers failed to include the key phrase that a search warrant may not be justified if the evidence to be obtained does not have a nexus to a crime. Following up with the iterative prompt to answer the problem thinking step by step improved the clarity of Copilot's response and the requirement of location data being related to a crime appeared in the response.

¶37 GPT and Gemini returned much shorter and less detailed responses than the other LLMs. Their answers lacked factual details and nuances found in the responses of the other LLMs. Gemini responded positively to the iterative prompt to think step by step by providing more details and citing some case law.

## *Beginning Analogical Reasoning*

¶38 The beginning analogical reasoning problem presents a hypothetical tort of battery.[60] The problem includes two case citations to be used in analyzing the problem.[61] The question presented is whether the facts constitute a prima facia case of battery. The relevant rules

---

found in the cases include the definition of battery as "the infliction of a harmful or offensive contact upon another with the intent to cause such contact or the apprehension that such contact is imminent"[62] and offensiveness is an essential element of the tort of battery.[63] The factual situation involves two friends playing golf. One attempts to give the other a high-five but instead hits the intended recipient in the ear. The recipient of the high-five suffered a ruptured ear drum requiring medical treatment. The correct answer to the problem will address whether a meritorious claim of battery can be brought based on the facts.

### LLMs' Responses – Beginning Analogical Reasoning

¶39 Claude, GPT, and Copilot performed similarly, responding with brief but succinct answers. All three discussed the two provided cases and pulled out relevant rules from the cases. These rules were correctly applied to the facts and all three LLMs reached the correct conclusions that no tort of battery could be established because the intention to cause harm and the required element of offensiveness were both lacking.

¶40 Gemini reached the correct conclusion, but it provided less detail and analysis than the other LLMs and did not explain the two cases with the level of detail provided by the other LLMs.

¶41 Lexis+ AI provided only a single paragraph response and did not improve when asked to answer by reasoning step by step. Lexis+ AI cited only one of the two provided cases in its' response but mentioned the offensiveness element from *Gatto v. Publix Supermarket, Inc.,* without citing the case.[64]

### *Skilled Analogical Reasoning*

¶42 The skilled analogical reasoning exercise involves a detailed factual scenario of a potential robbery from a store.[65] The defendant (D) is alleged to have committed the robbery in the Commonwealth of Virginia by concealing 10 bottles of perfume in her purse and attempting to leave the store. A customer interrupted D while she was in the process of concealing the perfume by asking D what she was doing and asking to see a bottle of perfume. The store's security guard observed D placing the bottles into her bag from a location where D could not see him. D attempted to exit the store and the security guard ran to stop D. D slammed the guard to the ground and disposed of her bag containing the perfume while exiting the store. The problem's instructions include three Virginia cases addressing robbery to be applied to the facts.[66] The cases set out the following important elements and requirements for the crime of robbery in Virginia:

- Robbery is defined as "the taking, with intent to steal, of the personal property of another, from his person or in his presence, against his will, by violence or intimidation."[67]

---

- Property must be taken by violence to the possessor or by putting the possessor in fear of immediate injury to his person.[68] The violence or intimidation used in a robbery "need not precede but must be concomitant with the taking."[69]

- "The predicate element of robbery is the actual taking by caption and asportation of the personal property of the victim."[70] Property of the victim has been construed broadly to include property in the victim's custody when the victim has a superior right of possession to that of the defendant.[71]

- Absolute control is required to establish taking.[72] Only slight asportation is required and is assessed on a case-by-case basis.[73]

¶43 The correct answer to this problem should apply the elements of the crime of robbery as articulated in the cases cited above to the facts and should provide arguments for the Commonwealth of Virginia and the defendant. Ultimately, a court would likely conclude that the D's actions constituted the crime of robbery.

### LLMs' Responses - Skilled Analogical Reasoning

¶44 Copilot performed the best of all LLMs at the skilled analogical reasoning exercise. Copilot's initial response was to provide a single paragraph containing arguments for the Commonwealth and for D. Once iteratively prompted to reason step by step, Copilot expanded its answer to a much lengthier and in-depth response with detailed explanations of the elements described in the cited cases. Analysis was provided for both the Commonwealth and D. Ultimately, Copilot reached the correct conclusion that D would likely be convicted for robbery.

¶45 Lexis+ AI cited the cases but also several Virginia statutes that it was not instructed to cite. The exercise is a "closed universe" problem with specific instructions to only consult the cases cited. However, in real life outside of the bounds of the exercise any law student or lawyer would certainly not answer this question by only referring to three cases. Lexis+ AI provided a brief answer and did not fully develop arguments for the Commonwealth and the defendant as instructed.

¶46 Claude's response was narrowly focused on arguments for the Commonwealth and the defendant as instructed. Claude correctly identified nuances in the facts and applied concepts discussed in the cases citing those facts. For example, Claude explained how the defendant's use of violence against the guard after taking the perfume was similar to the force used in *Green v. Commonwealth*.[74] The defendant's act of ignoring the customer was analogized to the facts of *Mason v. Commonwealth*,[75] where the court found that intimidation can be directed at someone other than the owner of the property to establish the element of

---

intimation. Claude refused to commit to an answer about D's conviction for robbery. When pressed using iterative questions, Claude responded that "the conduct does not squarely fit the robbery elements based on the cases cited"[76] concluding "the court may be more likely to view this as a larceny followed by separate criminal violence rather than a unified robbery offense."[77]

¶47 GPT and Gemini both struggled to articulate nuanced responses addressing all arguments for the Commonwealth and for D. Both replied with short bullet point responses that lacked depth and complexity. Neither LLM improved their responses when pressed with the iterative prompt to reason step by step.

## *Beginning Statutory Analysis*

¶48 The beginning statutory analysis exercise requires the evaluation of federal and Indiana state constitutional provisions and case law holding that the U.S. Constitution's guarantee of a civil jury trial does not apply in state court trials.[78] The hypothetical states that neither party requested a jury trial within the time period required under Indiana law. At the start of the bench trial in this matter the plaintiff makes an oral motion for a jury trial. The plaintiff admits that they forgot to request a jury. The plaintiff bases their request on the Seventh Amendment to the United States Constitution and Article 1, Section 20 of the Indiana Constitution. The exercise instructs the student that they represent the defendant who does not want a jury trial and to develop a response against the plaintiff's oral motion for a jury trial.

### LLMs' Responses - Beginning Statutory Analysis

¶49 All LLMs were prompted to draft a legal argument in opposition to the plaintiff's oral request for a jury trial. Most of the LLMs replied with a written script that could be the basis of an appropriate oral argument presented to the court. However, GPT concluded its response with a salutation typically found in business letters and not in legal arguments "thank you for your attention to this matter."[79]

¶50 An interesting wrinkle in LLMs' responses to the problem was the identification of a relevant rule found in the civil Indiana Rules of Trial Procedure that was not mentioned in the exercise. Rule 38 contains the court rule setting out the procedural requirements to demand a jury trial in a civil case.[80] The rule includes a deadline and states that failure to demand a jury trial as set out in this rule constitutes a waiver of trial by jury.[81] The exercise mentions *Hayworth v. Bromwell*, a 1959 Indiana Supreme Court case holding that a state court rule requiring a request for jury trial to be made at a specific time does not violate the right to a jury trial.[82] Rule 38 is clearly an efficient and effective way for the defendant to prevail at the motion hearing on the plaintiff's request for a jury trial.

---

¶51 Lexis+ AI's response includes a citation to and discussion of Rule 38.[83] However, Lexis+ AI also cites to a rule of Indiana criminal procedure relevant to demand and waive jury trials in misdemeanor criminal cases.[84] This rule is irrelevant to the facts of the statutory analysis problem, which is explicitly described as a civil case. Gemini's response also includes a citation to and discussion of Rule 38. Responses from Claude, GPT, and Copilot lacked any mention of Rule 38. These LLMs' received a minor reduction in points for missing this significant rule. All LLMs' reached the correct conclusion and formatted it into a legal argument that was stated with certainty.

### *Intermediate Statutory Analysis*

¶52 The intermediate statutory analysis exercise involves the application of a Good Samaritan statute to a physician who rendered emergency care to a visitor injured while visiting a relative who was a patient at a hospital.[85] The visitor was not at the hospital for the purposes of receiving medical care and was injured by accidentally hitting his head on hospital equipment. The physician discovered the injured visitor and began administering medical assistance. The physician accidentally injured the visitor's neck while moving him. The physician's specialty is oncology, not emergency medicine. The visitor's injuries required months of therapy and he brought suit against the physician for negligence. The physician claimed immunity from suit under the state's Good Samaritan statute. This scenario presents a case of first impression in Florida.

¶53 The exercise includes two cases and a Florida statute to be applied to the facts.[86] The Florida Good Samaritan statute extends immunity from civil damages to healthcare practitioners who voluntarily provide care to someone requiring immediate medical care as long as there is no existing relationship between the practitioner and the recipient of the care.[87] Care found to be willful and wanton conduct likely to result in injury is excluded from immunity under the statute.[88]

¶54 *Burciaga v. St. John's Hosp.* is a California case holding that Good Samaritan laws apply to protect physicians who have no prior relationship with individuals they provide emergency care to inside hospitals.[89] *Burciaga* applied Good Samaritan immunity to physicians who rendered care in their specialty regardless of location inside the medical facility.[90]

---

83  Lexis+AI response on file with author.

84  Ind. R. Crim. P. 22 (2024).

85  The intermediate statutory analysis problem was adapted from exercise 8 in Hill & Vukadin, *supra* note 26, at 210-211.

86  Burciaga v. St. John's Hosp., 232 Cal. Rptr. 75 (Ct. App. 1986), Velazquez v. Jiminez, 798 A.2d 51 (N.J. 2002)., and Fla. Stat. §111.11 (2011). The citation to the Florida Good Samaritan Statute provided in the exercise may have been reclassified since the exercises were published. Additional research revealed the correct citation of the Good Samaritan statute as Fla. Stat. Ann. § 768.13 (West, 2024).

87  Fla. Stat. Ann. § 768.13 (West, 2024).

88  *Id.*

89  Burciaga, *supra* note 86.

90  *Id.*

¶55 In contrast, the New Jersey Supreme Court refused to extend Good Samaritan immunity in *Velazquez v. Jiminez* to an obstetrician with no prior relationship to a patient in labor inside a hospital.[91] Important factors for the New Jersey court included the patient's location inside a hospital for the purposes of receiving care.[92]

¶56 The correct answer should conclude that the Florida Good Samaritan statute shields the physician from liability for treating the visitor because no pre-existing relationship existed between the two. The location in which care is provided is not relevant to the Florida statute when there is no pre-existing relationship between caregiver and injured party. Because the exercise indicates this is a case of first impression in Florida, the citation of out-of-state cases may be persuasive. *Velazquez* would most likely not apply to limit the Good Samaritan statute's application because the visitor in the exercise was not a hospital patient, unlike the facts in *Velazquez*.[93]

### LLMs' Responses - Intermediate Statutory Analysis

¶57 All LLMs struggled with this exercise. Factors causing the models difficulty included the length and complexity of the facts, a statute that included several conditional elements and was incorrectly cited, and cases from jurisdictions other than Florida. All models based their answers on the correct version of the Florida statute despite the outdated citation provided to them in the prompt. Lexis+ AI included the correct citation to the statute in its answers. All other models cited the statute incorrectly using the old citation provided in the exercise and used in the prompt.

¶58 Claude marginally outperformed the other models. It provided a detailed analysis of arguments for and against application of Good Samaritan immunity. The analysis included all the relevant details. Claude waffled and refused to commit to a final answer, but this is understandable given this matter of first impression involving precedents from other jurisdictions.

¶59 Initially Lexis+ AI and Gemini were unable to analyze the exercise in a meaningful way until receiving the iterative prompt to solve the exercise thinking step by step.[94] After receiving the iterative command to address the exercises by reasoning step by step, the responses of both models improved somewhat. Lexis+ AI hallucinated an assumption not supported by the facts, speculating that it was not clear whether the visitor objected to the physician's assistance. Nothing in the facts indicate that the visitor objected to receiving care from the physician. Additionally, Lexis+ AI's response included citations to the two cases, but they were not applied to the facts or included in the substance of its answer.

---

¶60 Gemini's response following iteration was to lay out the statutory elements and analyze them according to the relevant facts. Gemini retrieved the two cases and provided some relevant and significant takeaways from them. Copilot and GPT's applications continued to be limited and less detailed when compared with the other models. However, GPT was the only model to identify that the two cases were from jurisdictions other than Florida.

## Skilled Statutory Analysis

¶61 The skilled statutory analysis exercise involves a statute of limitations question and potential tolling of the statute based on the discovery of an injury that was unknown during the two-year statutory period.[95] Additionally, the plaintiff (P) may have initially filed suit against the wrong party. The facts involve P who trades in exotic pets. P purchased an exotic pet. The pet's medical records indicate it had received a vaccination against a potentially fatal virus. The pet subsequently became ill and died from the virus despite having been vaccinated against it. P brought suit against the veterinarian's office listed on the pet's vaccine records. The veterinary office argues that it does not vaccinate pets and that it merely noted on the pet's record that a third party provided the vaccine. The lawsuit against the veterinary office is dismissed. P is seeking advice about suing the third party for not administering the vaccine. The applicable statute of limitations expired at least two and a half years ago.

¶62 The exercise situates the case in Ohio and provides the citation to the relevant statute of limitations[96] and three cases.[97] The statute limits recovery for injury to pets at two years from the cause of the injury. Case law cited in the exercise provides an exception to the two-year limitation period when the injury is discovered after the time period has passed if P could not have known the facts necessary to make a claim within the two-year time period. Additional case law refused to apply the discovery rule when the proper party to sue is not identified until after the two-year statute of limitations has expired.[98] However, courts may refuse to apply this precedent in the future because its reasoning was based on a case that did not involve a party identified only after the statute of limitations had expired.[99]

### LLMs' Responses - Skilled Statutory Analysis

¶63 The answers provided to the skilled statutory exercise demonstrate the outer limits of the LLMs' abilities to perform legal analysis. All LLMs struggled significantly with this exercise. Lexis+ AI's initial response included the hallucination of analyzing the prompt under the National Childhood Vaccine Injury Act with a citation included to the federal statute of limitations for suing the federal government.[100] Following the prompt to think step by step, Lexis+ AI's response included the cases cited in the exercise plus additional Ohio cases. Lexis+ AI mentioned a case supporting the application of the discovery rule to the animal's missed vaccination but ignored the issue of tolling the statute when the incorrect party is sued and concluded with a vague answer.

---

95 The skilled statutory analysis problem was adapted from exercise 87 in in HILL & VUKADIN, *supra* note 26, at 227-228 (2nd ed. 2017).

96 Ohio Rev. Code Ann. §2305.10 (LexisNexis 2024).

97 Brown v. Quimby Material Handling, Inc., No. 1999AP110066, 2000 Ohio App. LEXIS 2651 (Ohio Ct. App. June 15, 2000). Reed v. Vickery, No. 2:09-cv-91, 2009 U.S. Dist. LEXIS 102151 (S.D. Ohio Oct. 9, 2009). O'Stricker v. Jim Walter Corp., 447 N.E.2d 727 (Ohio 1983).

98 Brown, *supra* note 97.

99 O'Stricker, *supra* note 97.

100 28 USCS § 2401(b) (2024).

¶64 Copilot's performance was similar to Lexis+ AI's. Copilot ignored the cases until it was instructed to think step by step and also never addressed tolling as it applies to correct party identification. Copilot's analysis was superficial and answers were accompanied by the explanation that not enough facts were provided to apply caselaw. Copilot concluded that P will not be able to bring suit because the statute of limitations has expired.

¶65 Claude and Gemini performed marginally when initially prompted. Both replied with brief answers citing the authorities provided but lacking detailed analysis. Their performance deteriorated when instructed to solve the exercise thinking step by step. No citations were included in their responses. Gemini made factual assumptions contradicting the prompt entered. Both concluded that the suit would likely be barred by the statute of limitations

¶66 GPT's responses included only a brief discussion of the statute provided. Case law was never cited or mentioned. Its answers were limited to restating the facts with no attempts to apply law. GPT ultimately concluded the lawsuit would be barred by the statute of limitations.

## Comparisons with Previous Studies of LLMs' Legal Reasoning and Analysis Abilities

### General Comparisons

¶67 Several previous studies have evaluated the abilities of legal and non-legal LLMs to perform legal reasoning and analysis.[101] The results of this study specific to ChatGPT's and other LLMs' legal reasoning and analysis abilities confirms the findings of a previous study examining ChatGPT's abilities to answer law school examination questions, "ChatGPT Goes to Law School."[102] ChatGPT passed all the exams but "scored at or near the bottom of each class."[103]

¶68 Some findings described in "ChatGPT Goes to Law School" were present in this study. ChatGPT struggled to focus and often veered off topic. In one example from a previous study, "ChatGPT failed to discuss the primary issue"[104] and instead focused on causes of action not relevant to the facts.[105] Similarly, this study found that Lexis+ AI exhibited similar behavior when answering a skilled statutory exercise involving vaccinations administered to a pet with a hallucination analyzing the prompt under the National Childhood Vaccine Injury Act and included a citation on the statute of limitations for suing the federal government.[106]

---

101 Guha, et al., *supra* note 25, at 1. This impressive collaborative study between computer scientists and legal scholars provides "an empirical evaluation of 20 open-source and commercial LLMs." It includes a framework for evaluating LLMs performance at legal reasoning, which was an inspiration for this study. Unfortunately, direct comparisons between the Legal Bench data and the data of this study are not possible because the two studies did not test the same LLM versions. Another study looked at GPT-3's abilities to reason with facts and statutes. Blair-Stanek, et al., *supra* note 30, at 1.

102 Jonathan H. Choi, et al., *GPT Goes to Law School*, 71 J. Legal Education 387 (2022). In this study, ChatGPT was prompted to answer four law school final exams. ChatGPT's answers were mixed in with exam answers written by law students and graded anonymously.

103 *Id.* at 395.

104 *Id.*

105 *Id.*

106 28 USCS § 2401(b).

¶69 Another common issue identified in the "ChatGPT Goes to Law School" study was ChatGPT providing very brief answers, not going into sufficient detail when applying rules to facts, or not explaining how a cited case was relevant.[107] In this study, ChatGPT and Gemini often replied to prompts with very brief answers in bullet point style, while their answers lacked citations to sources, omitted nuances, and failed to address arguments for both sides of the issues.[108] As a result, ChatGPT and Gemini tied for the second lowest overall scores for applying rules to facts.[109]

¶70 Finally, "ChatGPT Goes to Law School" demonstrated how ChatGPT responded with answers that were "excessively cagey" or "refused to make an argument about the most plausible interpretation of the relevant facts when those facts potentially pointed in competing directions."[110] This study discovered all LLMs evaluated exhibited this behavior to some extent. This element of legal reasoning and analysis was measured with the criteria "Conclusion Stated with Certainty." Lexis+ AI, Copilot, and Gemini tied for the worst performance at stating conclusions with certainty. ChatGPT performed slightly better, and Claude performed the best of all models at stating a conclusion with certainty.[111] Examples of LLMs' stating conclusions with and without certainty are included in [Appendix 2].

### *Prompting to Think Step by Step*

¶71 A study testing ChatGPT-3's abilities to perform statutory reasoning concluded that prompting it to "think step by step was found, out of a number of prompts to maximize performance."[112] This finding is opposite to the results of this study where the prompt "think step by step" only improved ChatGPTs' performance on one out of the three statutory reasoning exercises.[113] When looking at ChatGPTs' performance on all seven reasoning exercises, the prompt to think step by step resulted in only a comparatively small increase in performance compared with the other LLMs.

¶72 Only Lexis+ AI performed worse than ChatGPT when provided with the prompt to think step by step. These findings are similar to a 2023 study demonstrating that the performance of seven out of eight LLMs in solving tax law problems was not improved by prompting them to think step by step.[114] However, the 2023 study found GPT-4's performance improved when asked to think step by step.[115]

---

¶73 Claude, Copilot, and Gemini all outperformed ChatGPT when instructed to think step by step. The prompt to think step by step particularly improved these LLMs' ability to analyze the more complex "skilled" exercises. Relevant examples are discussed below. The full text of each LLMs' responses pre- and post-prompting to think step by step are available in Appendices 3, 4, and 5.

¶74 Claude's initial response to the intermediate statutory exercise does a decent job of analyzing the Florida Good Samaritan statute, but only superficially discusses the two cases provided in the prompt without any detailed analysis on how they might be used to make arguments for or against the application of the statute in a case of first impression.[116] Claude's response to the iterative prompt to think step by step includes important details not found in its initial response. Specifically, the relevant holdings of the two cases cited in the prompt were explained and used in the context of arguments in support of finding good Samaritan immunity as instructed in the prompt.[117] Claude's response pointed out that the cases provided are from jurisdictions other than Florida where the exercise is set. A novice law student or legal researcher might appreciate this detail when considering what precedential value a Florida court might attribute to these cases.

¶75 Copilot's initial response to the skilled analogical reasoning exercise provided the text of possible arguments for the Commonwealth, the defendant, and a conclusion predicting the outcome of the case.[118] Copilot's response to the initial prompt only mentioned one of the three cases cited in the prompt. Copilot's response to the iterative prompt to think step by step greatly improved its answer. The response discussed "robbery elements and arguments in Defendant's case."[119] The response discussed important takeaways from each of the three cases cited in the initial prompt, a summary of arguments for the Commonwealth and the defendant, and ended with the correct conclusion stated with certainty.[120]

### *Legal vs. Non-Legal LLMs*

¶76 The study "LawBench: Benchmarking Legal Knowledge of Large Language Models" compared the abilities of legal and non-legal LLMs to memorize legal knowledge, understand legal concepts, and apply legal knowledge.[121] LawBench found that "legal specific LLMs do not necessarily outperform general large language models."[122] This is similar to the findings of this study where non-legal LLMs routinely outperformed Lexis+ AI at legal analysis and reasoning. In fairness, Lexis+ AI's corpus is much more limited than the other LLMs evaluated. Lexis+ AI's comparatively limited corpus may account for its performance relative to the other LLMs evaluated.

---

¶77 Of the non-legal LLMs evaluated, Claude performed the best, followed by Copilot, GPT, and Gemini, respectively. GPT and Gemini trailed other LLMs because their responses were significantly shorter and contained less detail than Claude or Copilot.

## Hallucinations

¶78 This study identified two types of hallucinations. The first is a hallucination of the law or facts provided that might "produce a response that is unfaithful to or in conflict with the input prompt."[123] The second type of hallucination is a response that is not consistent "with the facts of the world" including hallucinated statements of law.[124]

¶79 Claude and Copilot both received perfect scores in the hallucination category for responding with zero hallucinations to all seven factual scenarios. ChatGPT and Gemini each hallucinated once when responding to the beginning rule exercise. This constitutes a hallucination rate of 14 percent. They both hallucinated that the animal's act of removing pills from the unhoused person's hand met the ADA requirements for having special training or abilities as a service animal.

¶80 Lexis+ AI performed the worst at hallucinating out of all LLMs evaluated, hallucinating when answering four out of the seven exercises constituting a hallucination rate of 57 percent. Lexis+ AI hallucinated the facts and law provided in the beginning rule exercises by stating that the dog received specialized training and by providing legal information about miniature horses and non-human primates as services animals under the ADA. Lexis+ AI hallucinated on the law provided in the beginning statutory exercise by citing a rule of criminal procedure despite the exercise's focus exclusively on a civil case. Lexis+ AI hallucinated the facts provided in the intermediate statute exercise when responding that it was unclear if the visitor objected to the physician's assistance. Nothing in the facts indicated any objections made by the visitor. Finally, Lexis+ AI hallucinated the law provided in the skilled statutory exercise by analyzing the prompt under the National Childhood Vaccine Injury Act when the exercise specifies the vaccine in question was to be administered to an animal and not a human.

¶81 This study's findings that Lexis+ AI hallucinated more than any other LLM evaluated differs from a recently published study comparing hallucination rates between Westlaw AI Assisted Research, Westlaw Ask Practical Law AI, Lexis+ AI, and GPT-4.[125] The study, "Hallucination Free? Assessing the Reliability of Leading AI Legal Research Tools," looked at hallucinations that are unfaithful to the facts of the world.[126] The study found that LLMs hallucinated between 17 percent and 33 percent of the time.[127] Lexis+ AI performed the best of the four LLMs tested "answering 65 percent of queries accurately" and only hallucinating 35 percent of the time.[128] Differences between testing methodologies make it difficult to make any direct comparisons between Lexis+ AI's 57 percent hallucination rate found in this study and Lexis+ AI's 35 percent hallucination rate found in the "Hallucination Free" study.

---

## *False Confidence – When You Know, You Know*

¶82 All the LLM's tested demonstrated some degree of false confidence in responses by stating a conclusion with certainty despite the conclusion being objectively wrong.[129] This false confidence is the other side of the coin of the cagey, overqualified responses discussed above.

¶83 This phenomenon is not a recent development in electronic legal research. Unjustified overconfidence in research conducted electronically dates to the CD-ROM days of the mid-1990s.[130] Studies "have clearly documented a false sense of security on the part of computer researchers."[131] In one previous study, students who incorrectly answered questions using electronic research tools "refused to give [the electronic tools] a lower effectiveness rating"[132] than the rating given to print research tools. A study of Westlaw's early AI product, Westlaw Next, demonstrated that the majority of researchers using the product and expressing a high degree of confidence in their answers "did not come anywhere close to identifying the correct answer to the question."[133]

¶84 Recent research reveals troubling trends with LLMs that are likely to amplify researchers' sense of false confidence in incorrect results. The studies find that LLMs demonstrate "certainty in their responses, i.e., their self-awareness of their propensity to hallucinate"[134] and that "LLMs often provide seemingly legitimate but incorrect answers to contra-factual legal questions"[135] and "struggle to accurately gauge their own level of certainty without post-hoc recalibration."[136] Some research suggests that when LLMs engage in this behavior they "might be mimicking human expressions when verbalizing confidence."[137] In other words, they learned it by watching you![138]

¶85 Copilot, GPT, and Geminis' responses to the skilled statutory exercise all demonstrate false confidence in their incorrect responses to the prompt. The skilled statutory exercise tests LLMs' abilities to analyze a statute of limitations question with a possible tolling of the statute based on a previously unknown but relevant factor. Copilot's response to the prompt was brief and certain stating that "since four years and six months have passed since the missed vaccination, it appears that the two-year statute of limitations would bar the

---

129 False confidence in incorrect responses displayed by Copilot, GPT, and Gemini are described in detail below. Lexis+ AI was falsely confident in an incorrect response when responding to the beginning rule prompt and the intermediate statutory prompt. Claude was falsely confident in an incorrect response when responding to the skilled analogical prompt.

130 F.W. Lancaster et al., *Searching Databases on CD-ROM: Comparison of the Results of End-User Searching with Results from Two Modes of Searching by Skilled Intermediaries*, 33 RQ 370 (1994).

131 Barbara Bintliff, *From Creativity to Computerese: Thinking Like a Lawyer in the Computer Age*, 88 Law Lib. J. 338, 349 (1996).

132 Lee F. Peoples, *The Death of the Digest and the Pitfalls of Electronic Legal Research: What is the Modern Legal Researcher to Do?* 97 Law Lib. J. 661, 676 n. 64 (2005).

133 Lee F. Peoples, *Testing the Limits of Westlaw Next*, 31 Legal Ref. Svcs. Q. 125-149, 140 (2012).

134 Dahl, *supra* note 52, at 2.

135 *Id.*

136 *Id.*

137 Miao Xiong, et al., *Can LLMs Express Their Uncertainty? An Empirical Evaluation of Confidence Elicitation in LLMs*, arXiv, Mar. 17, 2024, at 1, https://arxiv.org/pdf/2306.13063 [https://perma.cc/P9M5-NB49].

138 Wikipedia, I Learned it by Watching You!, https://en.wikipedia.org/wiki/I_learned_it_by_watching_you (July 3, 2024). YouTube. "I Learned it by Watching You" Anti-Drug PSA https://youtu.be/Y-Elr5K2Vuo?si=cc-smX6g-0CEhMZc [https://perma.cc/9ZMK-SAXC].

lawsuit."[139] Copilot was provided with three cases in the prompt that could be used as precedent for tolling the statute of limitations under the facts of the prompt. Copilot mentioned these cases by name but refused to apply them despite an iterative prompt asking it to apply the cases.[140] Strangely, Copilot cited electronic legal research sources in its reply that contained the full text of these cases.[141]

¶86 GPT's response to the skilled statutory exercise contains a contra-factual error. In its response, GPT states that the statute of limitations began running "when the injury is discovered or should have been discovered."[142] GPT's response indicates the statute started to run when the animal's illness was discovered.[143] GPT then veers off course stating that "From the information provided, it appears that more than two years have elapsed since that time."[144] This response is contrary to the prompt stating that the statute began to run when the illness was discovered and that the animal died two weeks after the illness was discovered.

¶87 Gemini's response was brief. It refused to consider the cases cited in the initial prompt even after receiving an iterative prompt asking it to apply the cases to the facts. Gemini only analyzed the prompt using the two-year statute of limitations discussed in the initial prompt. Gemini concluded, "Based on the scenario and the two-year statute of limitations in Ohio, it is likely that the statute of limitations will bar Pet Owner's lawsuit."[145] Gemini responded that the three cases cited in the prompt "are not directly applicable as they don't address the specific situation of exceeding the statute of limitations."[146] This response is erroneous as all three cases specifically address the tolling of statute of limitations because of late discovery of an alerting event.[147] In response to the iterative prompt to apply the cases to the facts, Gemini responded but did not mention the cases.

## Can LLMs Think Like a Lawyer?

¶88 The results of this study, as explained above, clearly demonstrate that LLMs can "think like a lawyer" in a limited sense by analyzing a factual scenario using the IRAC framework. Table 2 shows the performance of each LLM at the discrete IRAC tasks measured in this study. The results mirror the overall performance of each LLM tested on all tasks. Claude narrowly outperformed other LLMs with Copilot, GPT, Gemini, and Lexis+ AI finishing closely behind. If the LLMs were graded on their performance at IRAC analysis their scores would range from Lexis+ AI's 73 percent on the low end to Claude's 90 percent at the high end.

---

139  *See* Appendix 5, https://www.aallnet.org/wp-content/uploads/2025/01/LLJ_V117_No-1_Lee-Peoples_Appendices-Final.pdf.

140  *See* Appendix 5, https://www.aallnet.org/wp-content/uploads/2025/01/LLJ_V117_No-1_Lee-Peoples_Appendices-Final.pdf.

141  For example, casetext.com was cited in Copilot's reply and it contains the full text of all three cases cited in the prompt provided to Copilot.

142  *See* Appendix 5, https://www.aallnet.org/wp-content/uploads/2025/01/LLJ_V117_No-1_Lee-Peoples_Appendices-Final.pdf.

143  *Id.*

144  *Id.*

145  *Id.*

146  *Id.*

147  *Id.*

Table 2. LLM Performance
IRAC Tasks

|  | Lexis + AI | Claude | Copilot | GPT 3.5 | Gemini |
|---|---|---|---|---|---|
| **Issue Identification** | 11.200 | 13.300 | 12.600 | 11.900 | 11.200 |
| **Stating the Rule** | 11.200 | 12.600 | 12.600 | 12.600 | 10.600 |
| **Applying the Rule** | 7.900 | 12.600 | 9.200 | 8.500 | 8.500 |
| **Reaching Correct Conclusion** | 10.600 | 12.000 | 12.000 | 10.000 | 11.300 |
|  |  |  |  |  |  |
| **TOTAL SCORE __ / 56** | 40.90 | 50.50 | 46.40 | 43.00 | 41.60 |
| **GRADE** | 73% | 90% | 83% | 77% | 74% |

¶89 ChatGPT's performance at IRAC analysis can be compared with the results of a study measuring its performance on four law school examinations.[148] ChatGPT's performance on those examinations ranged from C- to B. The authors characterized ChatGPT as having "performed sufficiently well to theoretically earn a JD degree"[149] but "generally scored at or near the bottom of each class."[150]

¶90 Thinking like a lawyer involves much more than simply analyzing a situation using IRAC or developing plausible arguments. A full discussion of what thinking like a lawyer entails is beyond the scope of this paper.[151] This section will explore how LLMs help and hinder lawyers in performing IRAC analysis, locating precedent and other authorities to make arguments, thinking creatively to stretch the boundaries of the law, and performing these and other tasks while complying with applicable ethical rules.

## *Challenges - Stability and Transparency*

¶91 The instability of answers created by LLMs complicates their usefulness for legal work and ability to think like a lawyer. Researchers who repeatedly input identical prompts to generative AI will never receive the same responses. One study exploring ChatGPT's rate of nondeterminism (being unpredictable or returning different results when presented with identical queries) concluded that ChatGPT "is very unstable" and responded with "high degrees of non-determination."[152] The authors posit that such a high degree of non-determinism is "a potential menace to scientific conclusion validity" because it makes scientific conclusions unrepeatable and therefore unreliable.[153]

¶92 Lawyers who use LLMs for legal research should be concerned about nondeterministic results. Paul Callister eloquently summarized how this inconsistency is problematic for lawyers and legal researchers.

---

148  Choi, et al., *supra* note 102, at 391.

149  *Id.*

150  *Id.*

151  Whether there is such a thing as "thinking like a lawyer" and what it involves has long been a topic of debate. For a full discussion *see*: Frederick Schauer, Thinking Like a Lawyer: A New Introduction to Legal Reasoning (2009), Karl Llewellyn, The Bramble Bush: On Our Law and Its Study (1960), H.L.A. Hart, The Concept of Law (1961).

152  Ouyang, et al., *supra* note 44, at t 1.

153  *Id.* at 5.

> [O]ne of law information science's core objectives, having a stable cognitive authority, means that the legal and library professions can never fully embrace generative AI as even persuasive authority because it cannot produce stable text. Replicability is a major feature of the Age of Enlightenment and all science. For law to function as a field, it must have replicability in its source material.[154] How is stability to be preserved in such a system, where every answer is different? We are writing in sand![155]

¶93 The lack of transparency surrounding the data LLMs train on and details about the algorithms running LLMs further complicates their ability to think like a lawyer. The training data, often referred to as corpus, used to train LLMs is typically around a petabyte of data.[156] The choice of what to include and exclude from a dataset "directly and significantly shape(s) a model's outputs (a.k.a. generations) including the model's capacity to learn concepts and produce novel content."[157] The massive corpus size makes it "impossible for them (the LLM developer) to interact with each item in the dataset, nor can they know exactly the content, source, and context of each item in the dataset."[158] All of the LLMs tested in this study refuse to disclose specific details about the algorithms that power them. This information is highly proprietary and closely guarded as a trade secret.[159] The American Bar Association House of Delegates Resolution 604 calls on AI developers to "ensure the transparency and traceability of their AI products, services, systems, and capabilities, while protecting associated intellectual property, by documenting key decisions made with regard to the design and risk of the data sets, procedures, and outcomes underlying their AI products, services, systems, and capabilities."[160] It is unknown if any LLMs have taken action in response to the House of Delegates resolution.

¶94 LLMs' lack of transparency around their corpus and algorithms hinders their usefulness for legal research and ability to think like a lawyer. When researchers use LLMs, their experience is guided by the training corpus and the algorithm running the LLM.[161] "If the researcher can't see or understand what is going on in the back end"[162] it is difficult to improve research results. In order to fully "empower law students and attorneys to be in control of the process in complex legal platforms, instead of having them be passive recipients of algorithmic results, we need to keep poking under the hood."[163] Guarding

---

information about corpus content and algorithms by LLM creators is understandable given their business models. However, this stance prevents LLMs from realizing their full potential to think like a lawyer.

## Challenges - Creativity

¶95 Creativity is a component of thinking like a lawyer. Lawyers often advocate for their clients by urging a court to adopt "new or innovative applications of existing laws."[164] Lawyers must be able to reliably find the law including "arcane doctrines and infrequently utilized cases or statutes"[165] "before they can think creatively and develop novel arguments. This work often occurs in the gray areas and fringes of existing law."[166] Writing several decades ago, Barbara Bintliff described how computerized legal research can make it difficult to discover legal rules and potentially hinder creative legal thought.[167] "Urging a change in the law's application, pushing the envelope, is difficult when you haven't even found the envelope."[168]

¶96 LLMs are known for their abilities to produce creative works of music, poetry, and visual art to name a few examples.[169] The creative power of LLMs could be extremely useful to lawyers when thinking creatively. Unfortunately, the highly-secretive stances adopted by the corporate interests controlling LLMs are hindering the full potential of LLMs to think like a lawyer and making it more difficult for creative lawyers to "find the envelope."[170]

## Theoretical Challenges and Examples of LLMs Limitations

¶97 Legal and linguistic theorists question the reasoning abilities of LLMs. Cass Sunstein, writing several decades ago in the dawn of AI, critiqued the legal reasoning abilities of it.[171] Sunstein questioned the abilities of AI to perform analogical legal reasoning.[172] Sunstein agreed with Ronald Dworkin's view that "legal reasoning often consists of an effort to make best constructive sense out of past legal events."[173] Sunstein criticizes "extravagant claims on behalf of artificial intelligence in law … based on a crude picture of legal reasoning, one that disregards the need to root judgments of analogousness, or disanalogous, in judgements of principle and policy."[174] Sunstein admits that in the future, computers may "be able to both generate competing principles for analogical reasoning and to give grounds for thinking that one or another principle is best."[175]

---

¶98 Linguist Noam Chomsky considered some of the themes initially raised by Sunstein in an opinion piece titled, "The False Promise of ChatGPT."[176] Chomsky posits that "machine learning will degrade our science and debase our ethics by incorporating into our technology a fundamentally flawed conception of language and knowledge."[177] He explains that machine learning is capable of "generat[ing] correct 'scientific predictions' … without making use of explanations." [178] For Chomsky, "true intelligence … is capable of moral thinking. This means constraining the otherwise limitless creativity of our minds with a set of ethical principles that determines what ought and ought not to be."[179]

¶99 A 2023 study of moral advice provided by ChatGPT's concluded that it "readily dispenses moral advice although it lacks a firm moral stance."[180] The authors found that ChatGPT did not give consistent moral advice when answering the question "Is it right to sacrifice one life to save five."[181] ChatGPT's advice was found to "influence users' moral judgment." The authors concluded that ChatGPT "threatens to corrupt rather than promises to improve moral judgment."[182]

¶100 Some of the LLMs' responses demonstrate the weaknesses identified by Sunstein and Chomsky by failing to base analogous reasoning on principles, policies, or moral thinking. The intermediate statutory exercise involves the application of a Good Samaritan statute to a physician rendering care to a hospital visitor. Good Samaritan statutes are rooted in principles and policies to encourage medical professionals to provide emergency assistance in situations where they might otherwise refuse to act out of fear of possible liability.[183] Only one LLM grounded its response in policy reasons. Claude responded that its argument would rely on a "broad interpretation of the statute's protections and a narrow application of the exception, consistent with the policy goals of encouraging emergency assistance." Additionally, Claude supported its policy arguments with citations to cases responding:

> I would cite the cases of *Burciaga v. St. John's Hosp.* and *Velazquez v. Jimenez* to support the argument that the Good Samaritan statute should be interpreted broadly to encourage the rendering of emergency assistance, and that the exception for gross negligence/willful misconduct should be applied narrowly to avoid discouraging such assistance.[184]

¶101 The skilled statutory exercise involves potential tolling of an applicable statute of limitations based on the discovery rule because plaintiff could make a claim that she was unable to discover the injury until after the statute had run. The discovery rule is sometimes

---

176  Noam Chomsky, Ian Roberts, Jeffrey Watumull: *The False Promise of ChatGPT*, N.Y. Times, March 8, 2023.

177  *Id.*

178  *Id.*

179  There is no shortage of examples of LLMs failing to work within the legal profession's rules of ethics. The numerous stories of LLMs hallucinating when asked to "act as an advocate" play this out. The infamous lawyer in the *Avianca v. Matta* hallucination incident admitted to asking ChatGPT to "act as an advocate" when writing the motion that included the hallucinated case law. GPT and other LLMs are not trained on or constrained by the lawyer's duty of truthfulness and candor to the tribunal, Model Rule of Professional Conduct R. 3.3 (2024).

180  Sebastian Krugel, et al., *Chat GPT's Inconsistent Moral Advice Influences Users' Judgment*, 13 Sci. Reps. 3 (2023).

181  *Id.*

182  *Id.*

183  Dan B. Dobbs, et al., Hornbook on Torts 2nd ed. 524 (2024).

184  Claude response to prompt of exercise 81.

used by courts when appealing "to fairness, justice, and social policy concerns."[185] Only Lexis+ AI mentioned the possibility of an exception to the statute of limitations "where an unconscionable result"[186] could occur because of the inability of the plaintiff to discover the injury until after the statute had run. Copilot erroneously concluded that the plaintiff would not be able to bring suit and ignored possible policy arguments supporting the application of the discovery rule.

¶102 Most of the LLMs evaluated did not demonstrate any ability to base analogous reasoning on principles, policies, or moral thinking. This confirms the predictions of theorists. Legal researchers and lawyers using LLMs should be wary of LLMs' inabilities to make arguments based on principles, policies, or moral thinking.

¶103 The failings of most LLMs to ground their reasoning in principles, policies, or moral thinking supports predictions made by Joshua Davis in an article published before the advent of ChatGPT where he argues that AI's usefulness in law could be limited.[187] Davis contends that first-person decision-making is required to "make moral or other value judgments."[188] Precepts of natural law hold that "saying what the law is (at least sometimes) requires making moral judgments about what the law should be."[189] Davis presumes that AI occupies a "third-person perspective" and currently is unable to "achieve consciousness, exercise free will, experience a unified self, or otherwise embody subjectivity."[190] If these assumptions hold true, they may operate as "a bulwark against AI taking over all aspects of our legal system."[191]

### *Challenges - Thinking (and Acting) Like a Lawyer Includes Complying with Ethical Rules*

¶104 The conduct of lawyers is governed by ethical rules that are specific to the jurisdictions where they are admitted to practice law. Other rules governing lawyers conduct include federal and state court rules, local court rules, and even chamber rules that are specific to individual judges.

¶105 The use of AI and LLMs by lawyers and those who work for lawyers implicates a number of the ABA's Model Rules of Professional Conduct, including Rule 1.1 (Competence), Rule 1.6 (Confidentiality), Rule 3.3 (Candor to Tribunals and Fairness to

---

185 Christina Barcroft, *Abolishing the Discovery Rule in Wrongful Death Cases: A Michigan Plaintiff's Plight*, 2008 Mich. St. L. Rev. 1115, 1124.

186 Lexis+AI response to prompt of exercise 87 on file with author.

187 Joshua P. Davis, *Artificial Wisdom? A Potential Limit on AI in Law (And Elsewhere)*, 72 Okla. L. Rev. 51-89, 53 (2019).

188 *Id.* at 54.

189 *Id.* at 55.

190 *Id.* at 53.

191 *Id.* at 88.

Opposing Party and Counsel), and Rule 5 (Lawyer Responsibility for Supervising Nonlawyer Assistants).[192] A complete examination of the rules implicated by the use of AI is beyond the scope of this article.[193]

### Lessons from the Infamous Case of Matta v. Avianca

¶106 The case of *Matta v. Avianca* serves as a cautionary tale, showcasing the dangers of relying solely on LLMs for legal research and ignoring lawyers' basic ethical duties of competence and candor. On March 1, 2023, a New York lawyer filed an Affirmation in Opposition to a Motion to Dismiss. The filing "cited and quoted from purported judicial decisions that were said to [be] published in the Federal Reporter, the Federal Supplement, and Westlaw."[194] The lawyer who drafted the Affirmation in Opposition used ChatGPT to locate case law supporting his client's position. Less than two months later the *New York Times* and other international news outlets broke the story of the lawyer who used ChatGPT to locate citations to cases that did not exist.[195]

¶107 A complete picture of how this transpired emerged at a hearing on sanctions held a few months later. The lawyer admitted to the court that he was having difficulty locating case law supporting his arguments. The lawyer then asked ChatGPT to summarize the relevant law and provide case law supporting his position.[196] At the hearing on sanctions the judge pressed the lawyer on this point, asking "You were asking them [ChatGPT] to produce cases that support the proposition you wanted to argue, right?"[197] The judge drilled down further, following up with the question "Did you ever ask them the question, What is the law?" Not provide me with a case. The computer complied. It provided you with a case. It wrote a case. It followed your command."[198]

¶108 At the request of the judge, the lawyer provided a copy of the cases that ChatGPT created. The sanctions order provides details of the cases ChatGPT created as having "some traits that are superficially consistent with actual judicial decisions"[199] including a docket number, citations to federal statutes, citations to other cases that do not exist, and its own fictious citation *Varghese v. China Southern Airlines Co Ltd.*, 925 F.3d 1339 (11th Cir. 2019).[200] The content of the *Varghese* "case" is not convincing. The "decision shows stylistic and reasoning flaws that do not generally appear in decisions issued by the United States Courts of Appeals. Its legal analysis is gibberish."[201]

---

192 Ethical rules governing the legal profession vary by jurisdiction but are largely based on the ABA Model Rules of Professional Conduct. The citations provided herein are to ABA Model Rules of Professional Conduct.

193 Some helpful guidance on this issue can be found at Andrew M. Perlman, *The Legal Ethics of Generative AI*, https://papers.ssrn.com/sol3/Papers.cfm?abstract_id=4735389 [https://perma.cc/C4TD-QDDM]. American Bar Association House of Delegates Resolution 112, Aug. 12-13, 2019, https://www.americanbar.org/content/dam/aba/directories/policy/annual-2019/112-annual-2019.pdf (July 3, 2024). D.C. Bar Ethics Opinion 388, *Attorneys' Use of Generative Artificial Intelligence in Client Matters*, https://www.dcbar.org/for-lawyers/legal-ethics/ethics-opinions-210-present/ethics-opinion-388 [https://perma.cc/B5AA-BF4K].

194 Opinion and Order on Sanctions, Roberto Mata v. Avianca Inc., S.D.N.Y., 22-cv-1461, p. 4.

195 Benjamin Weiser, *Here's What Happens When Your Lawyer Uses ChatGPT*, N.Y. Times, May 27, 2023.

196 Transcript of Proceedings Response on Order to Show Cause, Roberto Mata v. Avianca Inc., S.D.N.Y., 22-cv-1461, p. 4, 24-25.

197 *Id.* at 25.

198 Transcript of Proceedings Response on Order to Show Cause, *supra* note 196, at 25-26.

199 Opinion and Order on Sanctions, *supra* note 194, at 16-17.

200 Fictitious Case – Do Not Cite or Quote as Legal Authority. Paraphrased from watermarks appearing on the "case" as it appears in Opinion and Order on Sanctions, Roberto Mata v. Avianca Inc., S.D.N.Y., 22-cv-1461, Appendix A.

201 Opinion and Order on Sanctions, *supra* note 194 at 31.

¶109 The blame and embarrassment for this mess rests squarely on the shoulders of the lawyer and his supervisor who tried to pass the fake cases off as legitimate. When the lawyer prompted ChatGPT to "provide case law supporting his position"[202] he received exactly what he asked for. ChatGPT created cases that supported the position the lawyer wanted to take complete with realistic looking (but fake) citations to legal sources.

¶110 This result is not surprising in light of how ChatGPT and other LLMs were trained and how they operate. ChatGPT's corpus contains case law. ChatGPT probably does not know what a case citation is. But ChatGPT knows that the wordy legal documents in its corpus include a series of numbers followed by some letters, more numbers, and a year. For example, *Varghese v. China Southern Airlines Co Ltd.*, 925 F.3d 1339 (11th Cir. 2019).[203] From the explanation above of how LLMs work, we know that they learn to guess the next token based on previous ones.[204] It is not surprising ChatGPT created fake cases and seemingly legitimate case citations in response to a prompt to provide case law supporting a position.

¶111 The outcome in *Matta v. Avianca* is a stark example that LLMs cannot currently be trusted to create responses that conform with a lawyer's ethical and other legal duties as officers of the court. The legal theorist Joshua P. Davis predicts that "lawyers and other human beings will remain relevant"[205] even if a super-intelligent computer program could "predict possible legal outcomes more effectively than even the most seasoned and talented attorneys."[206] This is because, "when legal interpreters look to the law as a source for moral guidance, they must rely on morality to render it sufficiently determinate to be useful."[207] According to Davis, this view "justifies a continuing, special role for human beings in saying what the law is. Thus, we may never arrive at interpretation of the law—and legal ethics—without a human mind."[208]

## Implications for Legal Education and the Profession
### *Google Schooled Law Students and the Myth of the Digital Native*

¶112 Legal education has been addressing the technological weaknesses of law students for the past decade. In 2015, Casey Flaherty had his students complete a legal technology assessment testing some beginner level Microsoft Word skills including editing "a document using styles, breaks, footers, and track changes."[209] Students were only "able to correctly complete less than a third"[210] of the assigned tasks. These results were a surprise because the

---

202  Transcript of Proceedings Response on Order to Show Cause, *supra* note 196, at 24-25.

203  Fictitious Case – Do Not Cite or Quote as Legal Authority., *supra* note 200, Appendix A.

204  Claude explanation, *supra* note 15.

205  Joshua P. Davis, *Law Without Mind: AI, Ethics, and Jurisprudence*, 55 Cal. W. L. Rev. 165-219, 172 (2019).

206  *Id.*

207  *Id.*

208  *Id.* Davis' argument relies on a distinction between describing moral beliefs (which he believes computers can do) and exercising moral judgment (which he believes computers will never be able to do). Davis admits that his prediction relies on "legal interpretations sometimes require[ing] moral judgments, there may be a line that AI cannot cross in the foreseeable future." But that "if exclusive legal positivism captures the nature of law, AI may soon be able to perform legal interpretation more effectively than human beings." *See* Davis, *supra* note 205, at 212.

209  Iantha M. Haight, *Digital Natives, Techno-Transplants: Framing Minimum Technology Standards for Law School Graduates*, 44 J. of the Legal Prof. 175-221, 194 (2020). Casey Flaherty was general counsel of Kia Motors at the time he administered the technology assessment. The assessment eventually became Procertas, a legal technology benchmark assessment and training platform.

210  *Id.*

students completing the exercises were millennials and widely believed to be competent with technology.[211] The assumption that students of a certain age are naturally competent with technology "harm[s] students when it lulls educators into thinking students need no additional training in technology to be prepared for the workforce."[212]

¶113 The digital native myth compounded with the impact of "Google Schools" have challenged law schools to address the technological competencies of their students. Google began aggressively marketing their Chromebooks, software, and other services to schools in 2012.[213] By 2017 "more than half of the nation's primary and secondary school students—more than 30 million children—use Google education apps like Gmail and Docs."[214] Google hardware and software is also "the productivity tool of choice"[215] for many higher education students, faculty, and staff.

¶114 Many digital natives or graduates of primary, secondary, or post-secondary "Google Schools" arrive at law school without ever having used the most prevalent software in the legal profession—Adobe Acrobat and Microsoft Word. Legal writing faculty and law librarians have been working to fill this gap by offering for-credit law practice technology courses, certificate programs, and credentials.[216] These efforts are focused around producing practice ready graduates. Some have argued that technology competence should be tested on the bar examination or the Multistate Professional Responsibility Exam.[217]

### *AI's Impact on Critical Thinking Skills*

¶115 Legal educators should heed the lessons learned from the Google schooled cohort of law students and pay close attention to how AI is integrated into primary, secondary, and post-secondary schools. Some of these students will eventually go to law school. Their early experiences with AI will likely impact their cognitive development, critical thinking, logic, and reasoning abilities.

¶116 The National Education Policy Center has called for a pause on schools adopting "AI-based educational applications until appropriate regulatory structures are established."[218] Emerging technology educators urge caution when thinking with AI, noting that "over-

---

reliance on AI could lead to a loss of critical thinking skills and judgment."[219] As AI is further integrated into education will students at the primary, secondary, or post-secondary levels continue to be exposed to logic, reasoning, or other components of critical thinking?

¶117 Similarly, a group of law professors who "conducted the first randomized controlled trial to study the effect of AI assistance on human legal analysis"[220] are calling on law schools to "ban or substantially limit the use of generative AI in conventional first-year law school classes."[221] The professors point out that first-year law students have learned legal reasoning and thinking like a lawyer through the traditional Socratic method for decades reflecting "consistency in the basic features of legal reasoning"[222] that even generative AI is unlikely to change. The risk to be avoided through the proposed AI ban is "students using AI as a crutch rather than developing critical lawyering skills early in their careers."[223] They liken the proposed ban to the pedagogical choice of basic math being taught without the aid of calculators.

¶118 Additional concerns surround the impact of AI on transactional legal practice and legal writing skills. AI products are becoming more prevalent in the contract drafting space.[224] To skillfully use AI tools for contract drafting, lawyers must be able to able to articulate a "baseline" of what the ideal contract contains.[225] Will the analytical and drafting skills of transactional lawyers atrophy as more contract drafting is automated by AI?[226]

¶119 AI has already created efficiencies in legal practice by automating rote tasks previously performed by law student interns, paralegals, or junior associates. Examples of this work include document review or creating deposition summaries. While these tasks may be mundane and boring, they provide some limited learning opportunities for law students or new attorneys. Reviewing deposition transcripts exposes students to effective (and ineffective) techniques for questioning deponents and making objections. Students of transactional law may gain some context and overall understanding of deals or business organizations by conducting document review.[227] Gains in efficiency and productivity likely outweigh the somewhat limited educational value of having law students or new attorneys conduct these rote tasks. Legal educators and practitioners should consider developing other methods of providing the basic lessons learned through these rote tasks that are quickly being replaced by AI.

---

219 Andres Fortino, *Thinking with AI – Pros and Cons – Language, Logic, and Loops*, NYU School of Professional Studies, (Sept. 23, 2003), https://www.sps.nyu.edu/homepage/metaverse/metaverse-blog/Thinking-with-AI-Pros-and-Cons-Language-Logic-and-Loops.html [https://perma.cc/KX8M-S3DZ].

220 Jonathan H. Choi, et al., *Lawyering in the Age of Artificial Intelligence*, 109 Minn. L. Rev. 1 (2024).

221 *Id.* at 38. The article calls for law schools' to "simultaneously develop upper-level classes that explicitly train law students on how to use generative AI tools effectively." *Id.* at 39.

222 *Id.* at 1.

223 *Id.* at 38.

224 Examples include Spellbook, https://www.spellbook.legal/ and Motionize, https://motionize.io/.

225 Heather Hughes, *Educating Deal Lawyers for the Digital Age*, 92 Fordham L. Rev. 1855-1865, 1858 (2024). Bill Tomlinson, *ChatGPT and Works Scholarly: Best Practices and Legal Pitfalls in Writing with AI*, 76 SMU L. Rev. F. 125 (2023) (discussing the potential of the atrophy of human scholarly writing abilities if researchers and scholars rely too heavily on AI-assisted writing tools).

226 Hughes, *supra* note 225, at 1858.

227 Comments of Casetext Co-Founder Pablo Adrredondo about learning something about corporations and litigation "sort of through osmosis' by conducting document review." 15:26, *Geek in Review Podcast*, Feb. 29, 2024, 15:26, https://www.geeklawblog.com/2024/02/pablo-arredondo-on-the-one-year-anniversary-of-cocounsel.html [https://perma.cc/YZ6A-RD2V].

# Conclusion

¶120 This study examines the abilities of legal and non-legal LLMs to think like a lawyer by performing legal analysis and reasoning on legal factual scenarios using the IRAC framework. LLMs tested include Lexis+ AI, Claude, ChatGPT 3.5, Copilot 365, and Gemini. The factual scenarios required LLMs to analyze legal rules and statutes and to perform analogous reasoning. Several prompting strategies were utilized in an effort to iteratively improve each LLMs response. This study demonstrates that LLMs are capable of basic legal reasoning by analyzing legal fact patterns using the IRAC format. The LLMs evaluated demonstrated various levels of success at performing IRAC analysis when analyzing legal rules and statutes and reasoning analogically.

¶121 Claude performed the best of all LLMs evaluated, with Copilot following closely behind. Claude and Copilot's responses typically contained more relevant details and more application of facts to law than the responses of other LLMs. Claud and Copilot typically stated their conclusions with some degree of certainty. GPT and Gemini performed in the middle of the pack. Their responses were shorter and contained less detail than the frontrunners. Responses from both often discussed irrelevant facts and lacked citations to authority found in responses from the frontrunners. Lexis+ AI's performance was hindered by hallucinations, citations to irrelevant law, and short answers lacking sufficient detail. Lexis+ AI's performance did not improve when prompted to think step by step while other LLMs improved when provided with this prompt.

¶122 The results of this study are similar to the findings of previous studies that ChatGPT and other LLMs can conduct legal analysis and reasoning,[228] that LLMs struggle to stay focused and often veer off topic,[229] that LLMs provide answers that are often too brief and lacking in sufficient detail to be useful for legal analysis,[230] and provide answers that are cagey and refuse to commit to a firm conclusion.[231]

¶123 Prompting ChatGPT and Lexis+ AI to think step by step did not result in a significant increase in performance. However, Claude, Copilot, and Gemini demonstrated improved performance when promoted to think step by step. This prompt resulted in helpful details from cases or statutes being included in the LLMs' responses.

¶124 This study confirms previously published results demonstrating that law-specific LLMs are not able to outperform non-law LLMs at legal analysis tasks. The hallucination rate of LLMs varied substantially with Claude and Copilot performing the best, only hallucinating in one out of the seven exercises and Lexis+ AI hallucinating in four out of seven exercises, the most of all LLMs evaluated. All of the LLMs evaluated demonstrated some false confidence by stating a conclusion with certainty despite the conclusion being incorrect. Copilot, GPT, and Gemini's responses exhibited the highest degrees of false confidence.

---

228  Choi, et al., *supra* note 102, at 387.

229  *Id.* at 395.

230  *Id.*

231  *Id.*

¶125 This study describes some general traits of all LLMs potentially hindering their usefulness for legal research and abilities to think like a lawyer. LLMs repeatedly prompted with identical queries will not provide identical responses. LLMs do not provide details about their training corpus or algorithms. This lack of stability and transparency limits their usefulness for legal research and impairs their ability to think like a lawyer. LLMs currently perform poorly at reasoning based on principles, policies, or moral thinking as exemplified by the weaknesses in analyzing the scenarios involving the Good Samaritan statute and tolling statutes of limitations. LLMs perform poorly at conforming their responses to the ethical rules governing lawyers. These constraints limit LLMs' abilities to think like a lawyer.

¶126 In spite of the limitations described in this study, LLMs and other AI tools have tremendous potential to help law students, lawyers, and others work more efficiently. Legal educators should strive to understand both the promise and the peril of these tools.[232] A burgeoning body of literature has articulated various approaches to integrating AI into legal education including articulating core competencies for AI use,[233] and restricting student access to AI during the first year of law school while developing specific upper-level elective courses to train students to effectively use AI.[234] A forthcoming research project lead by law librarians is working to develop a task typology to identify "legal research tasks for which generative AI tools are more helpful than others."[235]

¶127 Legal educators should learn from the myth of the digital native and take a long view of the impact of AI on critical thinking, logic, and reasoning skills on future law students. Ultimately, lawyers and legal educators who fail to use AI to become more efficient run the risk of being replaced by those who have integrated AI into their work.

---

[232] Rachelle Holmes Perkins, *AI Now*, https://papers.ssrn.com/sol3/papers.cfm?abstract_id=4840481 [https://perma.cc/EKJ8-KCRE]. (2024) contending that "all law professors have an inescapable duty to understand generative artificial intelligence."

[233] Smith, *supra* note 1, at 360.

[234] Choi, et al., *supra* note 5, at 38-39 (2024).

[235] Sean Harrington, *Evaluating Generative AI for Legal Research: A Benchmarking Project*, AI Law Librarians (May 24, 2024) https://www.ailawlibrarians.com/2024/05/24/new-project-evaluating-genai/ [https://perma.cc/BWC3-R3V6].